\RequirePackage{fix-cm}
\documentclass[smallextended]{svjour3}       
\smartqed  
\usepackage{amsmath}
\usepackage{graphicx}
\graphicspath{ {./} }
\usepackage{subcaption}
\usepackage{algorithmic}
\usepackage{url}
\usepackage{algorithm2e}
\usepackage{pdfpages}
\usepackage{multirow}
\usepackage[utf8]{inputenc}

 \journalname{}
\begin{document}

\title{Economic Hubs and the Domination of Inter-Regional Ties in World City Networks}

\titlerunning{Economic Hubs and Inter-Regional Ties}        
\author{Mohammad Yousuf Mehmood \and  Syed Junaid Haqqani \and Faraz Zaidi \and C\'{e}line Rozenblat 
}

\institute{Mehmood, MY \at School of Mathematics and Computer Science, Institute of Business Administration, Pakistan.
\email{m.mehmood.23375@khi.iba.edu.pk}
\and
Haqqani, SJ \at School of Mathematics and Computer Science, Institute of Business Administration, Pakistan.
\email{s.haqqani.23371@khi.iba.edu.pk}
\and
Zaidi, F \at School of Administrative Studies, York University, Canada.
\email{faraz.a.zaidi@ieee.org}
\and
Rozenblat, C \at Institute of Geography and Sustainability, University of Lausanne, Switzerland.
\email{celine.rozenblat@unil.ch}
}

\date{Received: date / Accepted: date}

\maketitle

\begin{abstract}

Cities are widely considered the lifeblood of a nations economy housing the bulk of industries, commercial and trade activities, and employment opportunities. Within this economic context, multinational corporations play an important role in this economic development of cities in particular, and subsequently the countries and regions they belong to, in general. As multinational companies are spread throughout the world by virtue of ownership-subsidiary relationship, these ties create complex inter-dependent networks of cities that shape and define socio-economic status, as well as macro-regional influences impacting the world economy.

In this paper, we study these networks of cities formed as a result of ties between multinational firms. We analyze these networks using intra-regional, inter-regional and hybrid ties (conglomerate integration) as spatial motifs defined by geographic delineation of world's economic regions. We attempt to understand how global cities position themselves in spatial and economic geographies and how their ties promote regional integration along with global expansion for sustainable growth and economic development. We study these networks over four time periods from 2010 to 2019 and discover interesting trends and patterns. The most significant result is the domination of inter-regional motifs representing cross regional ties among cities rather than national and regional integration.





	
	

\end{abstract}

\section{Introduction}\label{sec::introduction}

Cities are considered the prime economic engines of growth for any country \cite{taylor15}. Whether it is New York City in the USA, or London in England, Paris in France, or Tokyo in Japan, they are all examples of cities that not only drive the economies of their nations, but also have a significant economic impact on their geographic and economic regions \cite{csomos17,rozenblat17,currid06,drennan96}. One of the major reasons attributed to the global dominance for these cities is the presence of multinational companies \cite{godfrey99,alderson10,buckley20}. Multinational companies (MNC) bring in direct foreign investments, advanced technologies and stronger management capabilities to drive sustainable growth and economic development for cities, nations and world regions \cite{zhu15,hussain19}. 

Take the US Economy as an example, it is the largest in the world. MNCs represent only 1\% of total American firms. Yet, MNCs' parent companies constitute 24\% of the private sector GDP, account for 26\% of private sector employee compensation and engage in more than half of all US exports 
\cite{kim19}. They also account for more than 40\% of US imports. The Brookings Institution in Washington D.C. recognizes MNCs as playing a significant role in shaping the global economy \cite{kim19}, highlighting that \textit{their transnational activities have transformed the nature of international trade, investments, and technology transfers in the era of globalization}. The OECD estimates that MNCs generate half of global exports, 28\% of the world’s GDP, and nearly a fourth of employment opportunities worldwide \cite{backer18}. 

The study of cities in the context of MNCs has attracted lot of research. Certain cities balloon in growth to become \textit{world cities} \cite{taylor05} which command a sizable portion of the world’s gross domestic product (GDP), commerce, and trade. Not only that, but these cities facilitate transfer of knowledge and services throughout the world, which leads to the cross-pollination of various disciplines, fields, and industries, leading to significant economic impact \cite{lammarino13,rozenblat21}. 

Apart from these positive economic impacts, MNCs also negatively effect cities in several ways. A significant economic impact is the tax avoidance and revenue loss for cities. MNCs often establish complex organizational structures to minimize their tax obligations leading to revenue loss for host cities \cite{palan10}. Presence of large number of MNCs also create a dependency on foreign investments leaving the cities economically vulnerable \cite{hansen11}. Another common problem is gentrification, where the presence of MNCs cause an increased demand for real estate, which results in displacement of lower-income residents as affordability becomes an issue \cite{sassen01,maeckelbergh12}. Other negative impacts include environmental degradation \cite{rondinelli00}, exploitation of labor \cite{sethi12}, and closure of local businesses \cite{blomstrom98}.

Cities host multinational companies (MNCs) which are connected to their subsidiaries all around the world establishing strong ties among cities. Mapping a network of these ties provides an excellent proxy of economic connectedness and many researchers have studied these networks and discovered interesting results \cite{kraetke14,wall11,alderson04,rozenblat17,hussain19,saleem23} . 


There are several ways to analyze these networks. One way is to look at the bilateral ties which provide a microscopic view of the bigger network. Another way to study these networks is at the macro level, where clusters of interconnected cities emerge often representing the world's larger economic regions such as Northern and Southern America, Western and Eastern Europe, the Middle East, the Caribbeans, South Asia and so on. These networks can also be explored at the meso scale where small groups of inter-connected cities can be analyzed over time to understand the local (and implicitly the global) interplay among cities. One way to study these networks at the meso scale is to use network motifs. Network motifs, or simply motifs are connectivity patterns that play the role of a building block for these complex networks of cities \cite{milo02}. This approach to study complex networks in general, has been used in many other domains such as biological networks, transportation networks, social networks, and networks of electrical power grids\cite{omidi09,lacroix06,iovanovici19,alon07,zaidi13b,yu20}. Network motifs that occur much more significantly than chance, can be used to analyze deliberate connectivity trends and patterns revealing interesting global and local geopolitical and socio-economic interests \cite{le04,adams22,keating17}.

In this paper, we are interested in analyzing the connectivity patterns of cities from the top 15 countries of the world\footnote{Top 15 countries are selected based on GDP from the year 2022 https://globalpeoservices.com/top-15-countries-by-gdp-in-2022/} and the network motifs they form with the cities they have economic ties, and are spatially distributed in different economic regions of the world. The goal of this study is to discover how regional co-operation play a role in the economic stability and growth of cities that form the global epicenter of economic activities.

The rest of the paper is organized as follows: In section \ref{sec::related}, we discuss the related work from the perspective of research in network motifs; and networks of cities. Section \ref{sec::data} describes the data set used for analysis and section \ref{sec::methodology} explains the detailed methodology employed for experimentation. Following the analysis, we discuss our findings in section \ref{sec::results} and conclude the paper in section \ref{sec::conclusions}.



\section{Related Work} \label{sec::related}

\subsection{Network Motifs}

Network motifs are defined as patterns of sub-graphs that occur with a significantly higher frequency in a network. These motifs essentially define how a network is formed and are often termed as the building blocks of a complex network \cite{milo02}. Researchers have long used network motifs to analyze, understand and predict network behaviors \cite{stone19}. Algorithms to detect frequent motifs in large complex networks have found applications in a variety of domains such as ecology \cite{stone92}, social networks \cite{topirceanu16,zaidi13b}, epidemiology \cite{holland76}, maritime networks \cite{ducruet12}, and microorganisms \cite{alon07}.

Motifs have been used in spatial networks to uncover patterns and key sub-structures. For example, Iovanovici et al. \cite{iovanovici19} used network motifs to help urban planners in public transportation networks. Topirceanu et al. \cite{topirceanu14} used motifs to study road networks in cities to identify optimal road distributions. Domingues et al. \cite{domingues22} analyzed street networks within cities using motifs to characterize neighborhoods. Jin et al. \cite{jin19} study the structural properties of air passenger transport system between cities in China. Temporal analysis can also be done via network motifs to discover hidden information. This methodology of temporal analysis has been used to analyse large scale smart card data studying travel patterns \cite{lei20}.

Several algorithms exist to detect recurrent network motifs that require minimal supervision. Discovery of large size motifs (motifs with many nodes) has been shown as a computationally hard task \cite{yu20}. Algorithms have been developed to enable efficient discovery of motifs, however, the subject of motif discovery is still open ended and requires more research to reach more optimal techniques as our ability to capture and generate large graphs grows with technological advancements. MFinder \cite{kashtan02}, FanMod \cite{yu20} and MODA \cite{omidi09} are some of the widely used algorithms to discover motifs.

In the context of this paper, we didn't require the detection of complex motifs as we primarily focused on pre-defined motifs (with number of nodes fixed to four) that were readily present in the network. Searching for larger motifs  resulted in detecting motifs that were rare, and searching for smaller motifs (such as dyads and triads) resulted in a huge number of motifs. In both these cases, we would have either over fitted (for larger size motifs) or under fitted (for smaller size motifs) the economic ties obscuring relevant and interesting connectivity patterns.  

\subsection{Networks of Cities}

Cities and the interconnected networks that they form have been the focus of countless studies from a variety of different perspectives. However, the methodologies employed to study these networks differ to a large extent \cite{scott15}. Networks formed between these cities often demonstrate the properties of complex interconnected systems that can be analyzed to glean urban development theories and principles that guide the modern geographic studies \cite{pflieger10}. 

For example, Taylor and Derudder analyze the networks of cities formed as a result of multinational firms\cite{taylor15}. They provide a comprehensive analysis of how world city networks result in globalization, and the overall network connectivity impact city to city relations, regional cooperation and their impact on global economic crisis. Derudder also study the hierarchical and regional patterns in these networks of cities in an attempt to understand how certain cities and regions dominate the world economy\cite{derudder03}. Joyez studied the network of French multinational firms and the global value chains and discover how strategic selection of location is an important factor in achieving organizational goals\cite{joyez19}. Hussain et al. study these networks and identify central cities and how complex interplay between cities lead to world dominance \cite{hussain19}. Saleem et al. anaylyze the topological changes in these networks to determine economically evolving cities at a global scale \cite{saleem23}. 

Apart from analyzing networks of cities formed by multinational cooperation, several other methods have been studied to analyze networks of cities. For example, Silva et al. analyze wire transfers between different cities to demonstrate the economic inter-dependence of cities on each other\cite{silva20}. Mahutga et al. analyzed the networks of cities in the context of globalization, and distribution of power in the world system of cities\cite{mahutga10}. Blumenfeld-Lieberthal studied systems of transportation (air and railways) using these networks as an indicator of economic activity between cities \cite{blumenfeld09}. Sassen analyzed how social structures emerge from networks of cities and reinforce the importance of world cities \cite{sassen11}. Ducruet et al. analyzed port cities by constructing a maritime network of trade routes and discover hidden sub-structures such as specialized and long-distance trading links \cite{ducruet12}. Diao studied networks of high-speed rail to find growth in fixed asset investments in China \cite{diao18}. Pan et al.\cite{pan18} study the global financing of Chinese firms using data from initial public offerings on the Hong Kong stock exchange and draw interesting conclusions on how networks of cities are shaped by the global financing activities of Chinese firms. Rozenblat \cite{rozenblat21} study the intercity and intracity networks of cities at various scales to understand how spatial distribution of global cities play a role in the economic development of regions through global expansion.

A number of researchers have explored networks of cities trying to understand the process of innovation, often using data from patents, and research publications. A comprehensive review of the literature is out of the scope of this study but we review a few recent contributions from this field of study. Bianchi et al. studied inter-city networks using patent data to identify that broker cities connecting Latin American cities to other cities of the world has a negative influence on patenting outcomes \cite{bianchi23}. Cao et al. studied co-publication networks among Chinese cities to analyze intra and interregional collaboration linkages and their ability to innovate \cite{cao22}. Fan et al. propose a model to measure innovation efficiency of Chinese cities from 2003 to 2016 using capital as input and, patents and scientific papers as output. Results show how the innovation efficiency varies from one region to the other \cite{fan20a}. Guan et al. used patent data in the field of alternative energy to construct networks of cities as well as countries. Exploiting the structural properties of networks at multiple level, their analysis results in discovering complex relationships between a city's position in the network and its output in the form a patents produced. Yao et al. studied networks of cities built using patent data from 2001 to 2016 in China to analyze extralocal interactions in cities. Their results demonstrate a significant impact of intercity networks on a city's ability to innovate \cite{yao20}.

\section{Dataset}\label{sec::data}

\begin{figure}[h]
\centering
\includegraphics[scale=1.0]{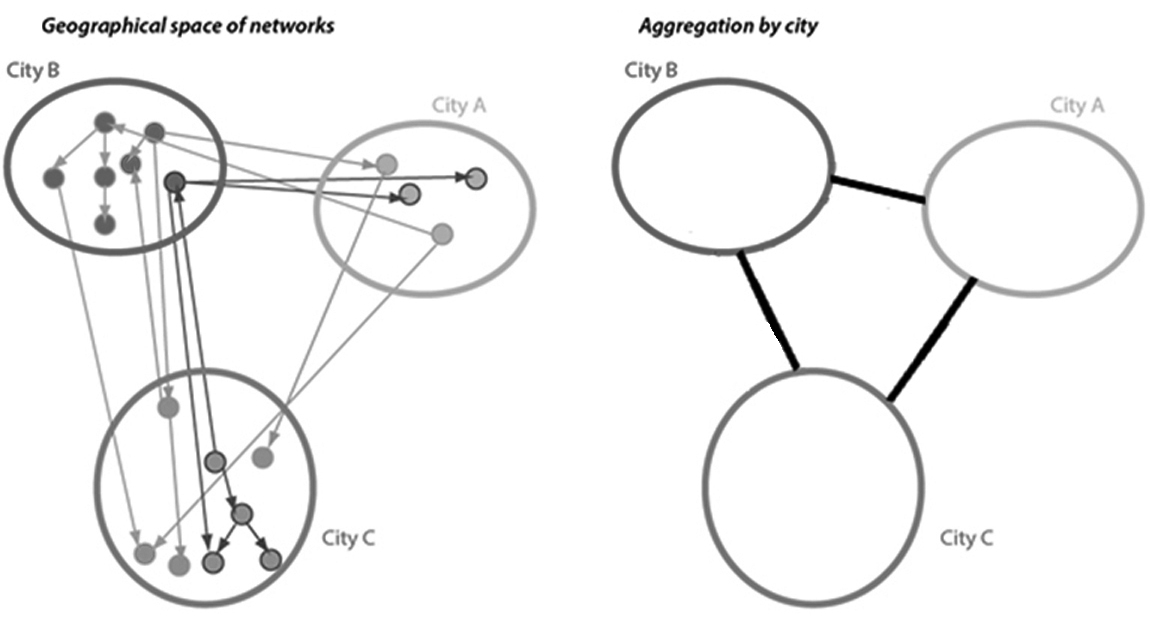}
\caption{Construction of Networks of Cities using Owner-Subsidiary Data from World's Top 3000 Group of Companies. (Left) Groups of Companies and their geospatial distribution in three cities; nodes represent companies and the links represent owner-subsidiary connections (Right) Aggregated Inter-city links resulting in an unweighted and undirected network.}
\label{fig::networkconstruction}
\end{figure}

The networks of cities based on economic ties studied in this paper are constructed using data from the world's top 3000 group of companies and their 800,000 direct and indirect subsidiaries located around the world. The top group of companies are selected based on their annual turnover. The data is made available by Bureau van Dijk\footnote{Source of Data : Bureau van Dijk Electronic Publishing (\url{http://www.bvdep.com/)} and was processed as a collaboration between the Universit\'{e} de Lausanne (CitaDyne group) and the University of Paris (ERC GeodiverCity group)}. The linkages between the direct and indirect subsidiaries are considered as a proxy for economic ties between the two cities that host the owner and subsidiary organization. The network thus constructed is converted to an undirected and unweighted network as we limit the scope of this research to analyze motifs for only undirected and unweighted networks. Using ties between cities based on multinational linkages has been studied extensively by other researchers such as \cite{kraetke14,wall11,alderson04} to represent economic networks and is considered to provide a good approximation accepted by urban geographers globally. 

Figure \ref{fig::networkconstruction} shows how these networks of cities have been constructed from the owner-subsidiary data. The figure explains the construction of the network using three groups of companies (Figure \ref{fig::networkconstruction}a) namely Group 1 (Pink), Group 2 (Green) and Group 3 (Blue) present in three different cities (namely City 1 (Yellow), City 2 (Red) and City 3 (Light Blue)). Starting from disconnected networks of groups of companies, two cities are connected to each other if they host two companies with owner-subsidiary link between them. Subsequently the links are aggregated over cities to obtain networks of cities. The data is available for four time periods: 2010, 2013, 2016 and 2019 which resulted in four networks being generated using the described procedure. 

The data set contains over 1500 cities located in over 190 countries and; encompassing 21 geographical sub-regions of the world as defined by United Nations Geoscheme\footnote{United Nations Geoscheme (https://unstats.un.org/unsd/methodology/m49/) divides the world into 6 regions, and 17 sub-regions.}. Figure 2 highlights the various sub-regions used in this analysis.

\begin{figure}[h]
\centering
\includegraphics[scale=1.4]{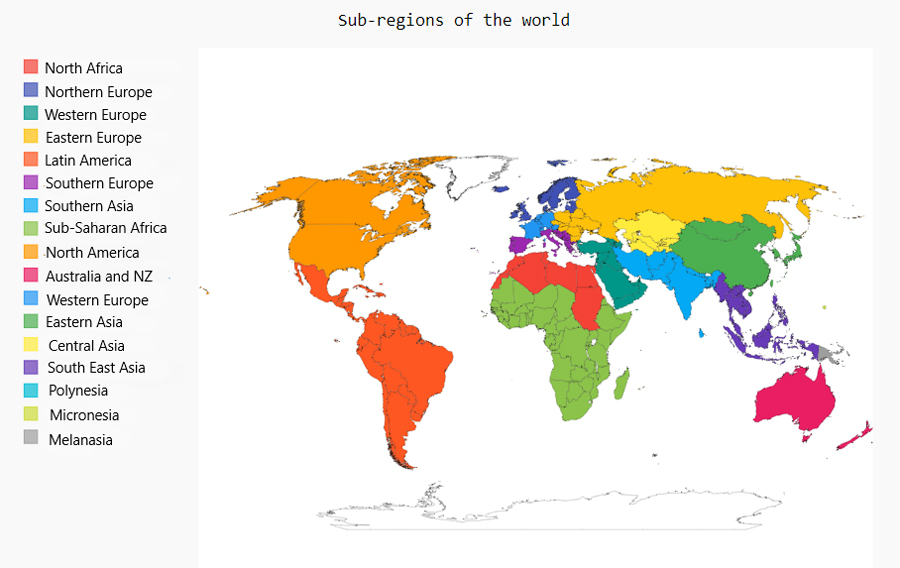}
\caption{Geographical sub-regions of the world as designated by United Nations Geoscheme.}
\label{fig::subregions}
\end{figure}

\begin{table}[h]
 	\centering
 	\caption{World's top 15 countries according to GDP in 2022 (Data From World Bank) and their cities with the highest number of cumulative connections with other cities over the studied period 2010 to 2019.}
\begin{tabular}{|l|l|l|l|}
 		\hline
 		\textbf{Rank} & \textbf{City} & \textbf{Country} & \textbf{Dollars} \\ \hline
 		1 & New York & United States of America & \$20.89 trillion \\ \hline
 		2 & Beijing & China & \$14.72 trillion \\ \hline
		3 & Tokyo  & Japan & \$5.06 trillion \\ \hline
 		4 & Munich & Germany & \$3.85 trillion \\ \hline
 		5 & London & United Kingdom & \$2.67 trillion \\ \hline
 		6 & Mumbai & India & \$2.66 trillion \\ \hline
 		7 & Paris & France & \$2.63 trillion \\ \hline
 		8 & Milan & Italy & \$1.89 trillion \\ \hline
 		9 & Toronto & Canada & \$1.64 trillion \\ \hline
 		10 & Seoul & South Korea & \$1.63 trillion \\ \hline
 		11 & Moscow & Russia & \$1.48 trillion \\ \hline
 		12 & Sao Paulo & Brazil & \$1.44 trillion \\ \hline
 		13 & Sydney & Australia & \$1.32 trillion \\ \hline
 		14 & Madrid & Spain & \$1.28 trillion \\ \hline
 		15 & Jakarta & Indonesia & \$1.05 trillion \\ \hline
 	\end{tabular}
 	\label{tbl::economichubs}
\end{table}

\begin{figure}[h]
\centering
\includegraphics[scale=0.2]{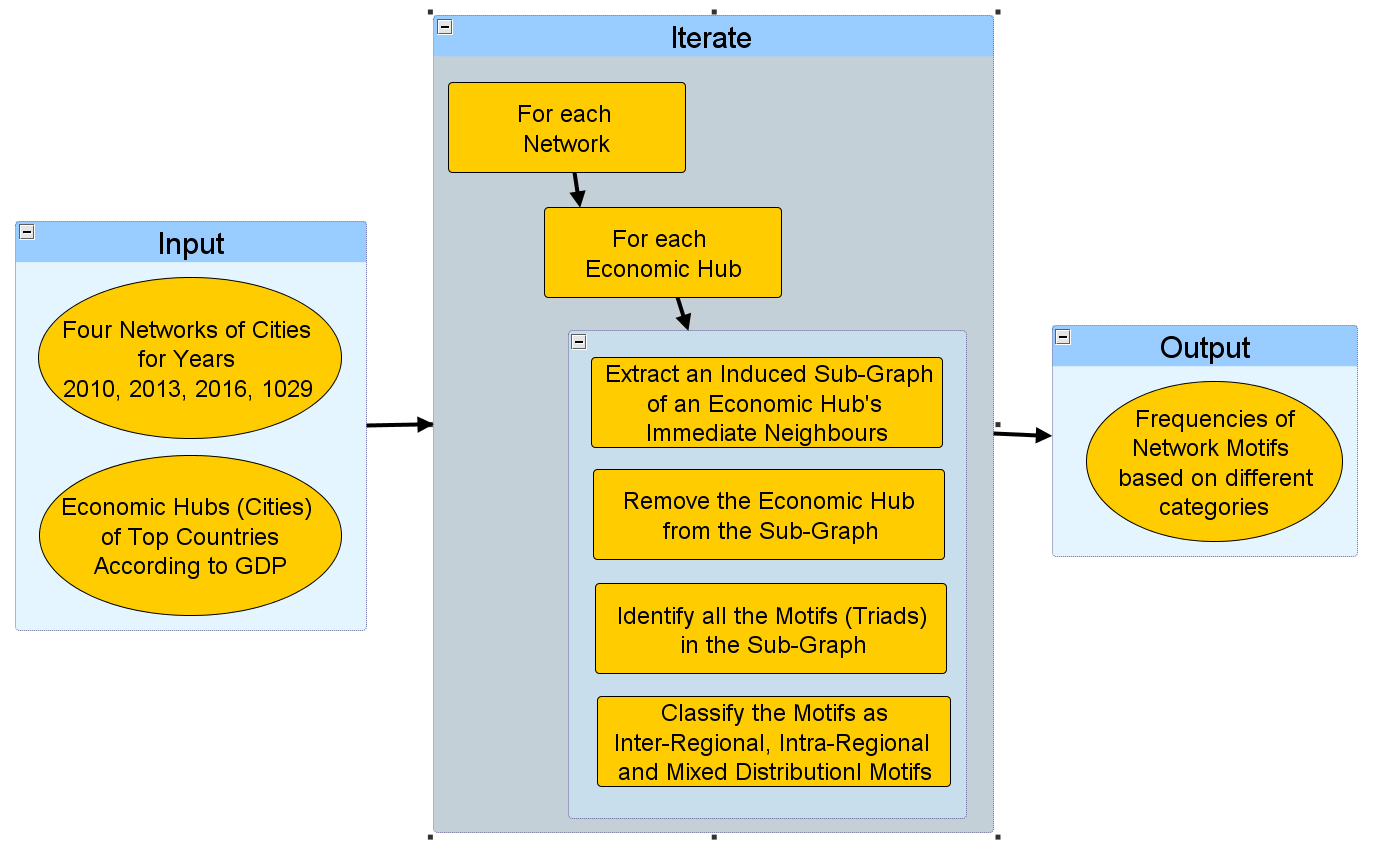}
\caption{Methodology used to find motifs in networks of cities for economic hubs (cities) selected from top countries based on GDP.}
\label{fig::methodology}
\end{figure}

\section{Methodology}\label{sec::methodology}

The overall methodology is depicted in Fig. \ref{fig::methodology}. Given the networks of economics ties over four periods (2010, 2013, 2016 and 2019), and global economic hubs identified as cities from the top 15 countries (as shown in Table \ref{tbl::economichubs}) based on GDP from the year 2022\footnote{https://globalpeoservices.com/top-15-countries-by-gdp-in-2022/}. One city from each of the 15 countries is selected for analysis. The city with the highest number of links in each of the 15 countries is selected as an economic hub representing their respective country. For example, in United States of America, the top three cities are New York, Wilmington (Delaware) and Chicago with 313,315, 139,976 and 101,302 links respectively summed over the four time periods from 2010 to 2019. Similarly, the top three cities for China are Beijing, Shanghai and Yichang; for United Kingdom, the top three cities are London, Birmingham and Edinburgh. We analyze network motifs discovered in the neighborhood of these selected cities and classify them into three categories: Inter-regional, Intra-regional and hybrid ties for each of the four years the networks are available. 

The motif identification algorithm works as follows: Based on each economic hub, an induced sub-graph of all its immediate neighbors is extracted. Then the economic hub is removed giving us a sub-graph which contains all the direct neighbors of the economic hub currently being processed. Next, all the triads from this sub-graph are extracted. As a result we are able to extract all the motifs of size four which are connected to each other and to the economic hub. The process is repeated for each economic hub for the four networks from 2010 to 2019. Once the motifs are extracted, they are classified into three categories which are discussed below:

\subsection{Motif Identification}

\begin{figure}[h]
\centering
\includegraphics[scale=0.8]{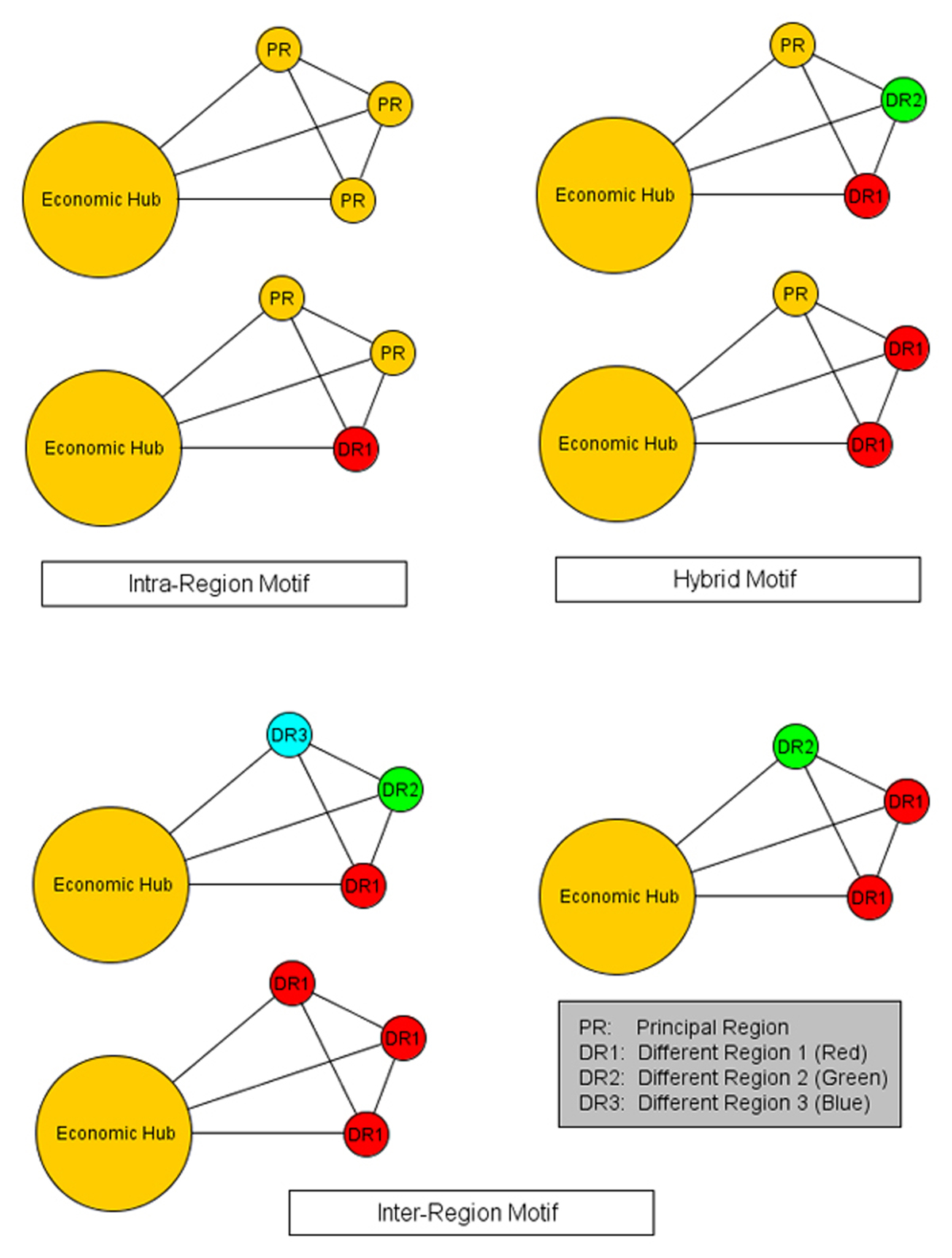}
\caption{Various configurations of Inter-region, Intra-region and Mixed Distribution motifs detected in networks of cities using economic sub-regions of the world.}
\label{fig::motifsdef}
\end{figure}

Based on the sub-regional distribution (as shown in Fig. \ref{fig::subregions}) we identify seven different configurations of motifs of size four where every node is connected to every other node. We use the following notation to explain the different configurations (as shown in Fig. \ref{fig::motifsdef}):

\begin{itemize}
\item Color Coding: The Economic Hub, PR, DR1, DR2, and DR3 are all cities. If they belong to the same sub-region, they are color coded with the same color
\item Economic Hub: represents one of the cities from the table \ref{tbl::economichubs} and is our city of interest.
\item PR: represents the principal sub-region where the city belongs to the same sub-region as the economic hub. PR cities always have the same color as the economic hub.
\item DR1, DR2, DR3: represents cities from different sub-regions and thus are color coded in different colors where DR1 is color coded red, DR2 is color coded green and DR3 is color coded blue.
\end{itemize}

These configurations are further grouped into three categories defined below:

\textbf{Intra-region Motifs:} Given a motif of four nodes, intra-region motifs are defined as cities predominantly belonging to the same region, or at most one city belonging to another sub-region as shown in Fig. \ref{fig::motifsdef}. The economic hub is represented by the big circle (yellow color) which is connected to three other cities. The economic hub is termed as the principal region of interest (denoted by PR). The two cases for the intra-region motifs are:
\begin{itemize}
\item The economic hub and the other three cities belong to the same sub-region and are represented by the same color
\item The economic hub and two other cities belong to the same sub-region whereas one city belongs to a different region denoted by DR1 (red color).
\end{itemize}

\textbf{Hybrid:} Given a motif of four nodes, hybrid motifs are defined as the case where the economic hub and one more city belong to the same sub-region, and the other two cities do not belong to the same sub-region as the economic hub. They may or may not belong to the same sub-region. The two cases for the mixed distribution motifs are:
\begin{itemize}
\item The economic hub and one other city belong to the same sub-region (represented by the same color) whereas DR1 and DR2 belong to two different sub-regions.
\item The economic hub and one other cities belong to the same sub-region whereas the other two cities belong to the same sub-region represented by DR1 (red color).
\end{itemize}

\textbf{Inter-region Motifs:} Given a motif of four nodes, inter-region motifs are defined as the case where the economic hub is connected to cities that do not belong to the same region as the economic hub. They may or may not belong to the same sub-region. The three cases for the inter-region motifs are:
\begin{itemize}
\item The economic hub is connected to three other cities each belonging to a different region denoted by DR1, DR2 and DR3.
\item The economic hub is connected to three other cities that all belong to the same sub-region as each other but to a different sub-region from the economic hub denoted by DR1.
\item The economic hub is connected to three other cities such that two of those cities belong to the same sub-region denoted by DR1 and one city belongs to another sub-region denoted by DR2. Note that all these three cities belong to a sub-region which is different from the sub-region of the economic hub.
\end{itemize}

The results are calculated based on these seven configurations grouped into three categories. The results are discussed in the following section.

\begin{table}[h]
 	\centering
 	\caption{Total number of motifs detected for the economic hubs for the years 2010, 2013, 2016 and 2019}
\begin{tabular}{|c|c|c|c|c|}
 		\hline
 		\textbf{City} & \textbf{2010} & \textbf{2013} & \textbf{2016} & \textbf{2019}\\ \hline
 		New York & 2316672 & 3438555 & 4828746 & 5541963 \\ \hline
 		Beijing & 141222 & 1024200 & 2626062 & 3565476 \\ \hline
		Tokyo  & 2752812 & 4261584 & 6039786 & 7177248 \\ \hline
 		Munich & 2060154 & 3065940 & 4790454 & 5072106 \\ \hline
 		London & 2662242 & 4140882 & 6030306 & 7078740 \\ \hline
 		Mumbai & 1080234 & 2034522 & 3956124 & 5765628 \\ \hline
 		Paris & 2293362 & 3433890 & 5139780 & 6088632 \\ \hline
 		Milan & 2120994 & 3574644 & 5387016 & 6115014 \\ \hline
 		Toronto & 1842168 & 2833746 & 4051398 & 4424892 \\ \hline
 		Seoul & 924792 & 1457160 & 2743566 & 5054370 \\ \hline
 		Moscow & 1078782 & 2051388 & 2775858 & 3670572 \\ \hline
 		Sao Paulo & 1158612 & 1848888 & 3564324 & 4044390 \\ \hline
 		Sydney & 1853286 & 3445128 & 5389176 & 5864154 \\ \hline
 		Madrid & 2218314 & 3646350 & 5148102 & 6213072 \\ \hline
 		Jakarta & 200514 & 421122 & 1305576 & 1679292 \\ \hline
 	\end{tabular}
 	\label{tbl::motifcount}
\end{table}

An important limitation of the proposed methodology is that as only one city is selected from each country, the analysis does leave out cities with higher number of overall links in the four networks. For example, Wilmington and Chicago which are the second and third ranked cities in the United States of America in terms of number of total links (from 2010 to 2019), have significantly more links than Seoul and Mumbai which are the leading cities from South Korea and India, but those cities are left out from this analysis. As mentioned earlier, the goal of this study is not to analyze cities with higher number of links (representing higher economic activity), but to analyze the connectivity patterns of cities from the leading countries based on GDP.

\subsection{Test of Statistical Significance}

Table \ref{tbl::motifcount} shows a clear tendency where inter-regional motifs dominate the hybrid motifs, and hybrid motifs dominate the intra regional motifs. We conducted an upper tailed pairwise t-test to ascertain this tendency for statistical significance across all cities and all four years. The results are presented in Table \ref{tbl::ttest}. Small p-values for all three pairwise test confirm that Inter regional motifs dominate hybrid and inter regional motifs, and hybrid motifs in turn dominate intra regional motifs.

\begin{table}[h]
 	\centering
 	\caption{P-Values to Test Statistical Significance to Ascertain the Domination of Inter Regional Links across all cities and all time periods}
\begin{tabular}{|c|c|}
 		\hline
 		\textbf{Pairwise Test} & \textbf{P-values}\\ \hline
 		Intra and Hybrid Regional Motifs 	& $2.36\text{e-}11$  \\ \hline
 		Intra and Inter Regional Motifs		& $1.78\text{e-}20$  \\ \hline
		Hybrid and Inter Regional Motifs  	& $2.91\text{e-}14$  \\ \hline 		
 	\end{tabular}
 	\label{tbl::ttest}
\end{table}

\section{Results and Discussion}\label{sec::results}

The total number of motifs of size four identified for each economic hub for the four years are listed in table \ref{tbl::motifcount}. There is a clear indication of an increasing trend for all the 15 cities listed in the table. With every increasing year, the number of connections show a clear positive trend whether it be trio of New York, Paris, London which are the major economic hubs of the world, or major European hubs such as Munich, Milan and Madrid. An interesting observation is the rapid exponential increase of Beijing from 2010 to 2013 and onward. China's economic growth in the past decade is well reflected in these numbers where the presence of multinational firms has significantly increased over the years. These are well reflected how Beijing is positioned as the economic hub of China, and the complex interplay of its ties with other cities of the world has increased Beijing's influence in the regional as well as global economy.

\begin{figure}
\centering
\includegraphics[scale=0.75]{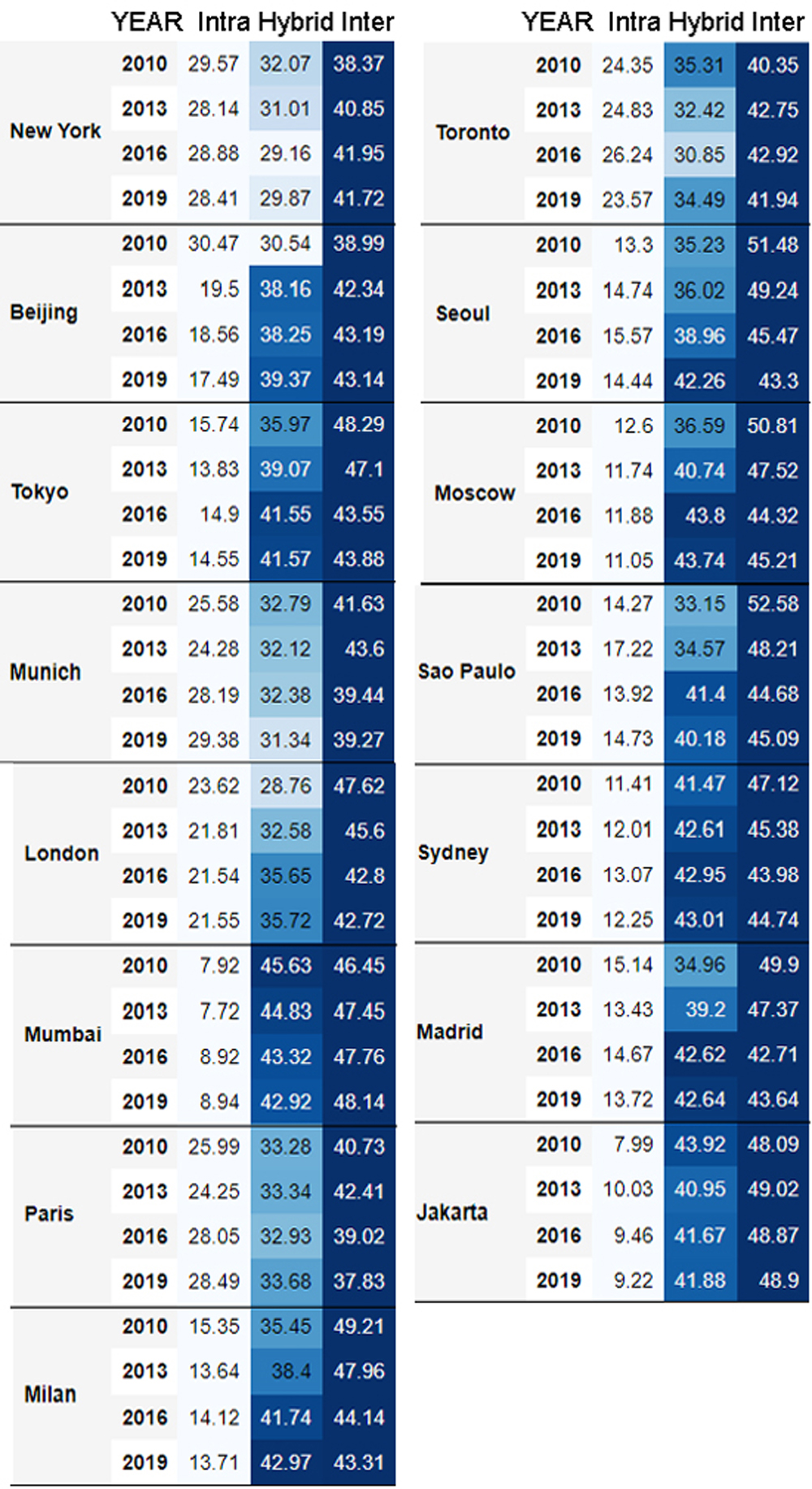}
\caption{World's Top Cities and the relative counts in terms of Inter-Region (first column), Hybrid Region (second column) and Intra-Region (third column) motifs. Color gradient indicates the comparative intensity where higher values are coded in dark blue and lower values are coded in white.}
\label{fig::hmap}
\end{figure}

The total count of motifs from various configurations based on sub-regional distribution are added up and the final counts are converted into percentages for easier comparison. The results are tabulated in Fig. \ref{fig::hmap} where each city is followed by the four years, and the percentages of Inter-Region, Intra-Region and Mixed Distribution motifs are shown for each of the four years. The table uses a color gradient from white (low values) to dark blue (high values). 

The most significant result of this research can be deduced from Fig. \ref{fig::hmap} which is consistently high values for Inter-Region motifs for all the cities across all four years. This is a clear indication of how economic ties formed by multinational firms are dominated by distant connections rather than having these ties with cities within the same country or the same economic sub-region. Stemming from the Internationalization Theory of multinational firms \cite{buckley20}, the proposed analysis provides an added quantitative analysis for reflection upon the expansion strategies implied by multinational firms. 

The economic hubs we have studied also appear as the world's most connected cities in the global city network of multinational firms \cite{sigler21} and interestingly, they all follow the same strategy with conscious and deliberate effort to succeed and establish global dominance. The selection of cities by multinational firms are generally motivated by several factors such as investment incentives, socio-political interplay, global business strategies \cite{goerzen13} but they all end up creating ties (owner-subsidiary relations of multinational firms) among cities that are inter-regional, located in distant economic regions. These distant ties come with their own set of challenges such as logistics and international trade agreements (or their lack of) but surprisingly enough, our study shows that the dynamics of economic geography are completely different from spatial geography where lands closer to each other do not necessarily dominate economic ties among cities.

Hybrid regional motifs appear second behind inter-regional motifs followed by intra-regional motifs and again, this phenomena can be observed for the all the cities studied in this paper. Although there are similarities, the ratio of the distribution of motifs varies from one city to the other. For example, world's largest economy represented by New York has $41.7 \%$ inter-regional motifs, $29.8 \%$ hybrid motifs and $28.4 \%$ intra-regional motifs in 2019. When compared with Jakarta which is $15th$ in the list of cities, the inter-regional motifs are nearly half of the total motifs $48.9 \%$ and along with hybrid motifs, make up around $90$ of the total motifs with only $9.2 \%$ intra-regional motifs indicating the low inclination towards national and within region economic ties. 

Moscow, Mumbai, and Sydney all exhibit a similar behavior where intra-regional ties make up around $10 \%$ of the total motifs and, inter-regional and hybrid motifs make the bulk of the motifs. Again this is an indication of strong distant economic alliances. Tokyo, Madrid, Sao Paulo, Seoul, Milan, and Beijing have around $80 \%$ to $85 \%$ motifs made up of inter-regional and mixed distribution motifs which is an indication of some national and sub-regional ties. Finally cities such as New York, Paris, Munich, London, and Toronto all have  fairly close distribution with values hovering around $30 \%$ for hybrid and intra-regional motifs. Cities such as New York and Toronto have strong regional ties in the North American region which is the reason why there is a higher tendency of within region ties, and equally likely are the hybrid motifs. Similar to this, Munich, London and Paris are influential cities in Europe, and demonstrate a fairly equal distribution between intra-regional and hybrid motifs.

\section{Conclusions and Future Work}\label{sec::conclusions}

In this paper we have analyzed networks of cities formed as a result of owner-subsidiary links of multinational firms using network motifs of size four. We focused on the world's top 15 cities and their ties formed with other cities in the context of inter-region, mixed region and intra-region motifs. Results reveal a strong inclination towards distant connectivity with inter-regional ties clearly dominating the other types of motifs. This phenomena was consistently observed for all the major economic hubs from around the world. The experiments also revealed significant growth in the total number of motifs formed by each city and again, this behavior was persistent across all the studied cities.  


Since this study is restricted to motifs of four nodes and the cities of the top 15 countries by GDP in 2022, this is just an initial step for further exploration whereby the types of motifs and cities under consideration can be increased to unearth valuable insights about the networks of multinational firms across world cities. Further, the analysis can be expanded to national, sub-regional and continental level to discover interesting connectivity patterns at various scales.

\bibliographystyle{plain}
\bibliography{VisuSnamMotif}

\begin{thebibliography}{10}

\bibitem{adams22}
Dawda Adams, Kweku Adams, Rexford Attah-Boakye, Subhan Ullah, Waymond Rodgers,
  and Danson Kimani.
\newblock Social and environmental practices and corporate financial
  performance of multinational corporations in emerging markets: Evidence from
  20 oil-rich african countries.
\newblock {\em Resources Policy}, 78:102756, 2022.

\bibitem{alderson04}
Arthur~S Alderson and Jason Beckfield.
\newblock Power and position in the world city system.
\newblock {\em American Journal of sociology}, 109(4):811--851, 2004.

\bibitem{alderson10}
Arthur~S Alderson, Jason Beckfield, and Jessica Sprague-Jones.
\newblock Intercity relations and globalisation: The evolution of the global
  urban hierarchy, 1981-2007.
\newblock {\em Urban Studies}, 47(9):1899--1923, 2010.

\bibitem{alon07}
Uri Alon.
\newblock Network motifs: theory and experimental approaches.
\newblock {\em Nature Reviews Genetics}, 8(6):450--461, 2007.

\bibitem{bianchi23}
Carlos Bianchi, Pablo Galaso, and Segio Palomeque.
\newblock The trade-offs of brokerage in inter-city innovation networks.
\newblock {\em Regional Studies}, 57(2):225--238, 2023.

\bibitem{blomstrom98}
Magnus Blomstr{\"o}m and Ari Kokko.
\newblock Multinational corporations and spillovers.
\newblock {\em Journal of Economic surveys}, 12(3):247--277, 1998.

\bibitem{blumenfeld09}
Efrat Blumenfeld-Lieberthal.
\newblock The topology of transportation networks: a comparison between
  different economies.
\newblock {\em Networks and Spatial Economics}, 9(3):427--458, 2009.

\bibitem{buckley20}
Peter~J Buckley and Mark Casson.
\newblock The internalization theory of the multinational enterprise: past,
  present and future.
\newblock {\em British Journal of Management}, 31(2):239--252, 2020.

\bibitem{cao22}
Zhan Cao, Ben Derudder, Liang Dai, and Zhenwei Peng.
\newblock Buzz-and-pipeline dynamics in chinese science: the impact of
  interurban collaboration linkages on cities innovation capacity.
\newblock {\em Regional Studies}, 56(2):290--306, 2022.

\bibitem{csomos17}
Gy{\"o}rgy Csom{\'o}s et~al.
\newblock Cities as command and control centres of the world economy: An
  empirical analysis, 2006-2015.
\newblock {\em Bulletin of Geography. Socio-economic Series}, 38(38):7--26,
  2017.

\bibitem{currid06}
Elizabeth Currid.
\newblock New york as a global creative hub: A competitive analysis of four
  theories on world cities.
\newblock {\em Economic Development Quarterly}, 20(4):330--350, 2006.

\bibitem{backer18}
Koen De~Backer and Sebastien Miroudot.
\newblock Multinational enterprises in the global economy. heavily debated but
  hardly measured.
\newblock {\em Policy Note. OECD Publishing Paris. Accessed January}, 15:2020,
  2018.

\bibitem{derudder03}
Derudder, Witlox, and Catalano.
\newblock Hierarchical tendencies and regional patterns in the world city
  network: a global urban analysis of 234 cities.
\newblock {\em Regional Studies}, 37(9):875--886, 2003.

\bibitem{diao18}
Mi~Diao.
\newblock Does growth follow the rail? the potential impact of high-speed rail
  on the economic geography of china.
\newblock {\em Transportation Research Part A: Policy and Practice},
  113:279--290, 2018.

\bibitem{domingues22}
Guilherme~S Domingues, Eric~K Tokuda, and Luciano da~F Costa.
\newblock City motifs as revealed by similarity between hierarchical features.
\newblock {\em arXiv preprint arXiv:2204.09104}, 2022.

\bibitem{drennan96}
Matthew~P Drennan.
\newblock The dominance of international finance by london, new york and tokyo.
\newblock {\em The global economy in transition}, pages 352--71, 1996.

\bibitem{ducruet12}
C\'{e}sar Ducruet and Faraz Zaidi.
\newblock Maritime constellations: a complex network approach to shipping and
  ports.
\newblock {\em Maritime Policy \& Management}, 39(2):151--168, 2012.

\bibitem{fan20a}
Fei Fan, Huan Lian, and Song Wang.
\newblock Can regional collaborative innovation improve innovation efficiency?
  an empirical study of chinese cities.
\newblock {\em Growth and Change}, 51(1):440--463, 2020.

\bibitem{godfrey99}
Brian~J Godfrey and Yu~Zhou.
\newblock Ranking world cities: multinational corporations and the global urban
  hierarchy.
\newblock {\em Urban Geography}, 20(3):268--281, 1999.

\bibitem{goerzen13}
Anthony Goerzen, Christian~Geisler Asmussen, and Bo~Bernhard Nielsen.
\newblock Global cities and multinational enterprise location strategy.
\newblock {\em Journal of international business studies}, 44(5):427--450,
  2013.

\bibitem{hansen11}
Michael~W Hansen and Eric Rugraff.
\newblock {\em Multinational corporations and local firms in emerging
  economies}.
\newblock Amsterdam University Press, 2011.

\bibitem{holland76}
Paul~W Holland and Samuel Leinhardt.
\newblock Local structure in social networks.
\newblock {\em Sociological methodology}, 7:1--45, 1976.

\bibitem{hussain19}
Owais~A Hussain, Faraz Zaidi, and C{\'e}line Rozenblat.
\newblock Analyzing diversity, strength and centrality of cities using networks
  of multinational firms.
\newblock {\em Networks and Spatial Economics}, 19(3):791--817, 2019.

\bibitem{iovanovici19}
Alexandru Iovanovici, Lilla Pellegrini, Anca-Maria Moscovici, and Monica Leba.
\newblock Network motifs uncovering hidden characteristics of urban public
  transportation.
\newblock In {\em 2019 IEEE 15th International Scientific Conference on
  Informatics}, pages 000143--000148. IEEE, 2019.

\bibitem{jin19}
Ying Jin, Ye~Wei, Chunliang Xiu, Wei Song, and Kaixian Yang.
\newblock Study on structural characteristics of china's passenger airline
  network based on network motifs analysis.
\newblock {\em Sustainability}, 11(9):2484, 2019.

\bibitem{joyez19}
Charlie Joyez.
\newblock {Alignment of Multinational Firms along Global Value Chains: A
  Network-based Perspective}.
\newblock GREDEG Working Papers 2019-05, Groupe de REcherche en Droit,
  Economie, Gestion (GREDEG CNRS), Universite Cote d'Azur, France, February
  2019.

\bibitem{kashtan02}
N~Kashtan, S~Itzkovitz, R~Milo, and U~Alon.
\newblock Mfinder tool guide.
\newblock Technical report, Weizman Institute of Science, 2002.

\bibitem{keating17}
Christine Keating and Tara Schmidt.
\newblock Opportunities and challenges for multinational corporations at the
  base of the pyramid.
\newblock {\em Sustainability Challenges and Solutions at the Base of the
  Pyramid}, pages 387--409, 2017.

\bibitem{kim19}
In~Song Kim and Helen~V Milner.
\newblock Multinational corporations and their influence through lobbying on
  foreign policy.
\newblock {\em Multinational Corporations in a Changing Global Economy}, 2019.

\bibitem{kraetke14}
Stefan Kraetke.
\newblock How manufacturing industries connect cities across the world:
  extending research on ‘multiple globalizations’.
\newblock {\em Global networks}, 14(2):121--147, 2014.

\bibitem{lacroix06}
V.~Lacroix, C.G. Fernandes, and M.-F. Sagot.
\newblock Motif search in graphs: Application to metabolic networks.
\newblock {\em Computational Biology and Bioinformatics, IEEE/ACM Transactions
  on}, 3(4):360--368, Oct.-Dec. 2006.

\bibitem{lammarino13}
Simona Lammarino and Philip McCann.
\newblock {\em Multinationals and Economic Geography}.
\newblock Edward Elgar Publishing Limited, Cheltenham, UK, 2013.

\bibitem{le04}
Philippe Le~Billon.
\newblock The geopolitical economy of resource wars.
\newblock {\em Geopolitics}, 9(1):1--28, 2004.

\bibitem{lei20}
Da~Lei, Xuewu Chen, Long Cheng, Lin Zhang, Satish~V Ukkusuri, and Frank Witlox.
\newblock Inferring temporal motifs for travel pattern analysis using large
  scale smart card data.
\newblock {\em Transportation Research Part C: Emerging Technologies},
  120:102810, 2020.

\bibitem{maeckelbergh12}
Marianne Maeckelbergh.
\newblock Mobilizing to stay put: Housing struggles in new york city.
\newblock {\em International Journal of Urban and Regional Research},
  36(4):655--673, 2012.

\bibitem{mahutga10}
Matthew~C Mahutga, Xiulian Ma, David~A Smith, and Michael Timberlake.
\newblock Economic globalisation and the structure of the world city system:
  the case of airline passenger data.
\newblock {\em Urban Studies}, 47(9):1925--1947, 2010.

\bibitem{milo02}
Ron Milo, Shai Shen-Orr, Shalev Itzkovitz, Nadav Kashtan, Dmitri Chklovskii,
  and Uri Alon.
\newblock Network motifs: simple building blocks of complex networks.
\newblock {\em Science}, 298(5594):824--827, 2002.

\bibitem{palan10}
Ronen Palan, Richard Murphy, and Christian Chavagneux.
\newblock {\em Tax havens: How globalization really works}.
\newblock Cornell University Press, 2010.

\bibitem{pan18}
Fenghua Pan, Ziyun He, Cheng Fang, Bofei Yang, and Jinshe Liang.
\newblock World city networks shaped by the global financing of chinese firms:
  A study based on initial public offerings of chinese firms on the hong kong
  stock exchange, 1999-2017.
\newblock {\em Networks and Spatial Economics}, 18(3):751--772, 2018.

\bibitem{pflieger10}
Geraldine Pflieger and Celine Rozenblat.
\newblock Introduction. urban networks and network theory: the city as the
  connector of multiple networks.
\newblock {\em Urban Studies}, 47(13):2723--35, 2010.

\bibitem{rondinelli00}
Dennis~A Rondinelli and Michael~A Berry.
\newblock Environmental citizenship in multinational corporations: social
  responsibility and sustainable development.
\newblock {\em European Management Journal}, 18(1):70--84, 2000.

\bibitem{rozenblat21}
Celine Rozenblat.
\newblock {\em Intracity and intercity networks of multinational firms,
  2010-2019}, chapter~25, pages 511--556.
\newblock Edward Elgar Publishing, 2021.

\bibitem{rozenblat17}
Celine Rozenblat, Faraz Zaidi, and Antoine Bellwald.
\newblock The multipolar regionalization of cities in multinational firms'
  networks.
\newblock {\em Global Networks}, 17(2):171--194, 2017.

\bibitem{omidi09}
A.~Masoudi-Nejad S.~Omidi, F.~Schreiber.
\newblock Moda: An efficient algorithm for network motif discovery in
  biological networks.
\newblock {\em Genes Genet. Syst.}, 84(5):385--395, 2009.

\bibitem{saleem23}
Mohammed~Adil Saleem, Faraz Zaidi, and C\'{e}line Rozenblat.
\newblock World city networks and multinational firms: An analysis of economic
  ties over a decade.
\newblock {\em Networks and Spatial Economics}, 1:1--20, 2023.

\bibitem{zaidi13b}
Arnaud Sallaberry, Faraz Zaidi, and Guy Melan\c{c}on.
\newblock Model for generating artificial social networks having community
  structures with small-world and scale-free properties.
\newblock {\em Social Network Analysis and Mining}, 3:597--609, 2013.

\bibitem{sassen01}
Saskia Sassen.
\newblock {\em The Global City: New York, London, Tokyo}, volume~f.
\newblock Princeton University Press, 2 edition, 2001.

\bibitem{sassen11}
Saskia Sassen.
\newblock {\em Cities in a world economy}.
\newblock Sage Publications, 2011.

\bibitem{scott15}
A.J. Scott and M.~Storper.
\newblock The nature of cities: the scope and limits of urban theory.
\newblock {\em International Journal of Urban and Regional Research},
  39(1):1--15, 2015.

\bibitem{sethi12}
S~Prakash Sethi.
\newblock {\em Multinational corporations and the impact of public advocacy on
  corporate strategy: Nestle and the infant formula controversy}, volume~6.
\newblock Springer Science \& Business Media, 2012.

\bibitem{sigler21}
Thomas Sigler, Kirsten Martinus, and Julia Loginova.
\newblock Socio-spatial relations observed in the global city network of firms.
\newblock {\em PLOS ONE}, 16:e0255461, 08 2021.

\bibitem{silva20}
Thiago~Christiano Silva, Diego~Raphael Amancio, and Benjamin~Miranda Tabak.
\newblock Modeling supply-chain networks with firm-to-firm wire transfers.
\newblock {\em Expert Systems with Applications}, 190:116162, 2022.

\bibitem{stone92}
Lewi Stone and Alan Roberts.
\newblock Competitive exclusion, or species aggregation?
\newblock {\em Oecologia}, 91(3):419--424, 1992.

\bibitem{stone19}
Lewi Stone, Daniel Simberloff, and Yael Artzy-Randrup.
\newblock Network motifs and their origins.
\newblock {\em PLoS computational biology}, 15(4):e1006749, 2019.

\bibitem{taylor15}
Peter Taylor and Ben Derudder.
\newblock {\em World city network: a global urban analysis}.
\newblock Routledge, 2015.

\bibitem{taylor05}
Peter~J Taylor.
\newblock Leading world cities: empirical evaluations of urban nodes in
  multiple networks.
\newblock {\em Urban studies}, 42(9):1593--1608, 2005.

\bibitem{topirceanu16}
Alexandru Topirceanu, Alexandra Duma, and Mihai Udrescu.
\newblock Uncovering the fingerprint of online social networks using a network
  motif based approach.
\newblock {\em Computer Communications}, 73:167--175, 2016.

\bibitem{topirceanu14}
Alexandru Topirceanu, Alexandru Iovanovici, Mihai Udrescu, and Mircea Vladutiu.
\newblock Social cities: Quality assessment of road infrastructures using a
  network motif approach.
\newblock In {\em 2014 18th International Conference on System Theory, Control
  and Computing (ICSTCC)}, pages 803--808. IEEE, 2014.

\bibitem{wall11}
Ronald~Sean Wall and GA~Van~der Knaap.
\newblock Sectoral differentiation and network structure within contemporary
  worldwide corporate networks.
\newblock {\em Economic Geography}, 87(3):267--308, 2011.

\bibitem{yao20}
Li~Yao, Jun Li, and Jian Li.
\newblock Urban innovation and intercity patent collaboration: A network
  analysis of china's national innovation system.
\newblock {\em Technological Forecasting and Social Change}, 160:120185, 2020.

\bibitem{yu20}
Shuo Yu, Yufan Feng, Da~Zhang, Hayat Bedru, Bo~Xu, and Feng Xia.
\newblock Motif discovery in networks: A survey.
\newblock {\em Computer Science Review}, 37, 2020.

\bibitem{zhu15}
T~Zhu, Yago Aranda~Larrey, and Valerie-Joy Santos.
\newblock What do multinational firms want from cities?
\newblock Technical report, World Bank, Washington, DC, 2015.

\end{thebibliography}

\end{document}